# Formation Wing-Beat Modulation (FWM): A Tool for Quantifying Bird Flocks Using Radar Micro-Doppler Signals

Jiangkun Gong, Jun Yan*, Deyong Kong, Ruizhi Chen, and Deren Li

*Abstract*— **Radar echoes from bird flocks contain modulation signals, which we find are produced by the flapping gaits of birds in the flock, resulting in a group of spectral peaks with similar amplitudes spaced at a specific interval. We call this the formation wing-beat modulation (FWM) effect. FWM signals are micro-Doppler modulated by flapping wings and are related to the bird number, wing-beat frequency, and flight phasing strategy. Our X-band radar data show that FWM signals exist in radar signals of a seagull flock, providing tools for quantifying the bird number and estimating the mean wingbeat rate of birds. This new finding could aid in research on the quantification of bird migration numbers and estimation of bird flight behavior in radar ornithology and aero-ecology.**

## I. INTRODUCTION

Birds' flapping wings can modulate the radar echoes from bird flocks, but the underlying scattering mechanism remains poorly understood. In 1939, R.M. Page first documented radar signals from birds and observed that the fading of the echo correlated with the bird's wing-beating [1]. Subsequent research also showed that the radar signals from birds sometimes fade abruptly, and this fading is correlated with the flapping of the bird's wings[2][3]. The signal fluctuation pattern produced by wing flapping contains information about the wing-beat frequency and flapping modes, which can be used to detect birds' radar signals from nonbirds and classify bird species. Recent studies have used modern radar systems to investigate radar echoes from flying birds in field tests and found that wings can significantly contribute to the overall radar cross-section (RCS) values of birds [4][5]. Bird RCS often exhibits 10 dB-level fluctuations, and the related radar reflectivity of bird flocks can fluctuate up to 20 dBz [6][7]. Despite some modified scattering models proposed by other scholars to describe the scattering characteristics of birds [8][9], the scattering mechanism modulated by wing-beats remains poorly understood.

In this study, we explore the modulation of radar echoes from bird flocks by the flapping of their wings. We introduce the formation wing-beat modulation (FWM) effect as a mechanism to explain this modulation in Section II. Through the analysis of radar echoes from a sea bird flock, we provide evidence to support our claims and discuss the potential application of FWM in radar signals of bird flocks in Section III. We conclude by highlighting the significance of our findings and suggesting future research directions.

## II. METHODS

**Theoretical model**

The flapping wings of birds can produce micro-Doppler signals in radar echoes. The differential velocity of each wing with respect to the radar causes a distributed Doppler shift around the main Doppler shift in the spectrum [1][10]. As shown in Fig. 1a, the micro-Doppler velocity of the wingtip can be expressed as a function of the flapping rate of the wing, effective wing length, and phase angle, which is given by equation (1).

$$v_{mD}(t) = \omega L cos(\varphi_t) \quad (1)$$

where $\omega$ = the flapping rate of the wing, $L$ = the effective wing length, and $\varphi_k$ = the phase angle, which is equal to $\varphi_t = \omega t + \varphi_0$, and $\varphi_0$ is the initial phase angle. According to the Doppler effect, the Doppler frequency is the function of Doppler velocity and the wavelength, which is given by,

$$f = \frac{-2v}{\lambda} \quad (2)$$

And then, the micro-Doppler frequency of the flapping wing is rewritten in,

$$f_{mD}(t) = \frac{-2L\omega}{\lambda} sin(\varphi_t) \quad (3)$$

Given, $\varphi_t = \omega t + \varphi_0$, and then the $f_{mD}(t)$ is also given by,

$$f_{mD}(t) = \frac{-2L\omega}{\lambda} sin(\omega t + \varphi_0) \quad (4)$$

---

Jiangkun Gong, Jun Yan, Ruizhi Chen, and Deren Li are with State Key State Key Laboratory of Information Engineering in Surveying, Mapping and Remote Sensing, Wuhan University, No. 129 Luoyu Road, Wuhan, China (e-mail: gjk@whu.edu.cn, yanjun_pla@whu.edu.cn, ruizhi.chen@whu.edu.cn, drli@whu.edu.cn. Deyong Kong is with School of Information and Communication Engineering, Hubei Economic University, No.8 Yangqiaohu Road, Wuhan, China (e-mail: kdykong@hbue.edu.cn)

*Corresponding author: Jun Yan (yanjun_pla@whu.edu.cn; +86-027-68778527)

where, $\lambda$ = the radar wavelength.

Based on the Doppler effect, the micro-Doppler frequency of the flapping wing can be calculated using equation (3) which is rewritten in equation (4) as a sinusoidal modulation on the phase angle. This signature can be observed on Time-Frequency distribution images, typically processed using Short Time Fourier Transform (STFT).

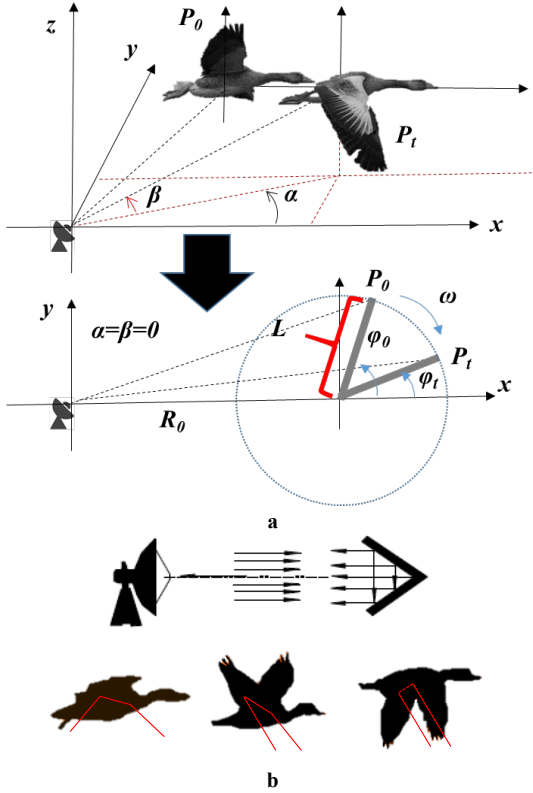

Fig. 1.  Structure of a bird wing, and the time-varying wingbeat corner reflector model. **(a)** the geometry of the radar and the flapping wing of birds, **(b)** the time-varying wingbeat corner reflector model.

The limited radar dwell time of bird detection radar systems may result in insufficient micro-Doppler data. Specifically, if the radar dwell time is denoted by $T_r$, the instantaneous micro-Doppler frequency can be calculated using the following formula:

$$f_{mD} = \int_0^{T_r} \frac{-2L\omega}{\lambda} sin(\omega t + \varphi_0) \, dt \tag{5}$$

$$f_{mD} = \frac{-2L}{\lambda}[cos(\omega t + \varphi_0) + c]\big|_0^{T_r} \tag{6}$$

where, $c$ = the integration constant. The equation (6) indicates that the Doppler frequency of a flapping wing is a constant in the spectrum of radar signals of birds. Given a flock of birds in one radar bin, the Doppler frequency of the flapping wing of the kth bird in the flock is

$$f_{k,mD} = \frac{-2L_k}{\lambda}[cos(\omega_k t + \varphi_{0,k}) + c_k]\big|_0^{T_r}, k = 1,2,3 \ldots N \tag{7}$$

where, $N$ = the bird number. Generally, the flock is composed of birds of the same species, the flapping rate, wing length, and body Doppler speed are usually identical, meaning that $L = L_k$ and $\omega = \omega_k$. The Doppler frequency of the flapping wing of the kth bird in the flock can be simplified as follows:

$$f_{k,mD} = \frac{-2L}{\lambda}[cos(\omega T_r + \varphi_{0,k}) - cos(\varphi_{0,k})], k = 1,2,3 \ldots N \tag{8}$$

The equation (8) states that the spectrum of radar echoes from bird flocks will exhibit multiple independent peaks. These peaks are modulated by the flapping wings in the flock, with the modulation period determined by the flapping rate ($\omega$) and the initial phase angle ($\varphi_{0,k}$) of the kth bird in the flock. This modulation is referred to as the "Formation Wing-Beats Modulation (FWM)" effect.

The amplitudes of the FWM peaks are determined by the flapping angle of each wing in the flock. In our earlier project, we proposed that the bird radar signals are modulated by the time-varying Wing-Beat Corner Reflector (WCR) effect[5], as depicted in Figure 4b. Using our WCR model, the Radar Cross Section (RCS) of the kth bird in the flock can be expressed as follows:

$$\sigma_k = \frac{8\pi a^2 b^2}{\lambda^2} \phi_k \tag{9}$$

where, $a$ = the length of the corner face; $b$ = the width of the corner face; $\lambda$ = the wavelength; and $\phi_k$ = the function relating flight gait to the radar observing angle of the kth bird. This amplitude can be obtained using Fourier transform, which is expressed as follows:

$$A_{f_{k,mD}} = \int_0^{T_r} \sigma_k e^{-j2\pi t} \, dt \tag{10}$$

Assuming that the radar observing the angle is neglected, and then $\phi_k$ is only equivalent to the initial phase angle, $\varphi_{0,k}$, and then equation (10) is simplified by

$$A_{f_{k,mD}} = \int_0^{T_r} (\frac{8\pi a^2 b^2}{\lambda^2} \varphi_{0,k}) e^{-j2\pi t} \, dt \tag{11}$$

The equation (11) demonstrates that the amplitudes of FWM peaks are directly proportional to the flapping angles, and each FWM peak has a unique amplitude that is not shared with other peaks.

The FWM signals exhibit a variable frequency spacing between adjacent peaks. Recent studies propose that birds in formation flight adopt a phasing strategy to manage the dynamic wakes created by their flapping wings [11], which is

$$\phi_s = \phi_t - 2\pi\lambda_f \tag{12}$$

where $\phi_s$ = the spatial phase, which is the function of the phase angle, of a bird in the flock, $\phi_t$ = the temporal phase, and $\lambda_f$ = the flight wavelength. Formula (12) suggests that the flapping gaits of a bird flock exhibit specific patterns in both time and space, albeit over a random distribution. A spatial phase of zero would theoretically indicate that the birds are flying in a direct following formation, resulting in matching wingtip paths. This would lead to patterned flight phases and a constant frequency spacing between adjacent FWM peaks. However, in practice, the spatial phase is not zero, implying that the flapping angles are not equal and causing a slight floating interval between FWM peaks.

**Application significance**

The number of FWM peaks corresponds to the number of birds in the flock. According to Formula (8), each FWM peak represents a single bird. Therefore, the bird count (N) in a radar bin can be determined using the FWM signals, as expressed below:

$$N = N_p - 1 \quad (13)$$

where $N_p$ = the peak number in the spectrum and 1 means the body peak. By computing the FWM signals from all radar bins, we can estimate the number of birds in the flock.

The FWM provides estimates for the wing-beat frequency and wingspan of bird species. Assuming that the birds are following each other directly, the spatial phase of the bird flock is zero, and the wingtip paths are aligned. As a result, the interval between FWM peaks becomes a constant value that is related to the wing-flapping rate and the number of birds. Therefore, FWM and Jet Engine Modulation (JEM) are similar in this respect. The group rate of flapping wings, with a mean flapping rate of $\bar{\bar{w}}$, can be viewed as a rotating rate of $N\bar{\bar{w}}$ in a rotating blade system.

$$W = N\bar{\bar{w}} \quad (14)$$

where $W$ = the rotating rate of blades. Since the spectral interval between the adjacent JEM peaks, $\Delta f$ is the function of the rotating rate of blades, which is [21]

$$\Delta f = NW \quad (15)$$

where $N$ = the number of blades. Thereby, the mean wing-flapping rate (i.e., $\bar{\bar{w}}$ with the unit of Hz,) of the bird species in the flock are calculated using FWM signals by, respectively,

$$\bar{\bar{w}} = \Delta f_{mD}/N^2 \quad (16)$$

Estimating wing-beat frequency in radar data of one range cell provides information for researching the bird species in a radar beam. Therefore, FWM may be applicable for identifying bird species based on radar signals in radar ornithology.

## III. RESULTS

**Simulated Results**

To demonstrate the FWM signals, we first use simulations. Assuming a radar dwell time of 20 ms and a wavelength of 0.03 m, we consider a bird's wing length of 0.6 m and a rotating rate of 7 Hz with a flapping angle range of -90° to 90°, sampled at intervals of 10. Using equation (8) (see Methods), we can determine the Doppler frequencies of the flapping wings of birds in the flock, which are shown in Fig. 2. There are 37 points in Fig. 2, each representing a bird with a specific phase angle, modulating the radar wave to produce different micro-Doppler frequencies. Even with a sampling interval of 5°, the intervals of micro-Doppler frequencies are not equidistant and are modulated by the sine function, as shown in Fig. 2b. However, there are always intervals, meaning that micro-Doppler frequencies related to the flapping wings in bird flocks can be detected and identified. Higher frequency resolution improves the detection of these micro-Doppler frequencies.

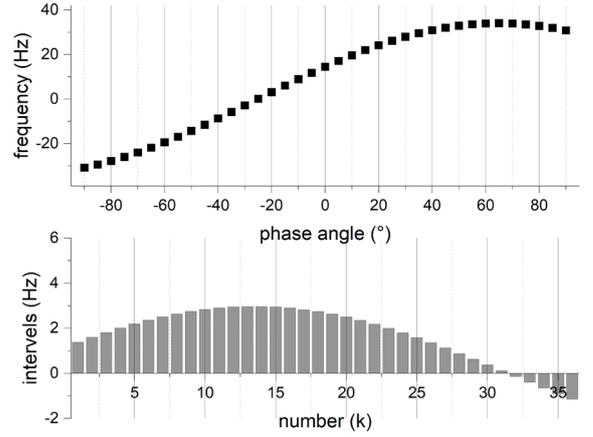

Fig. 2. Simulated micro-Doppler frequencies modulated by different flapping wings in a bird flock.

**Real Results**

We collected radar data using a pulse-Doppler phased array radar equipped with an active electronically scanned phased array antenna operating in the X-band, with a range resolution of 12 m and a Doppler velocity resolution of 0.3 m/s. The data was collected in an area near the mouth of the Yangtze River on the Yellow Sea coast of Nantong city, Jiangsu Province, China, over several days in October 2020. The sea class level was approximately 3, and we observed mainly seagulls and other sea birds, as it was not a migratory season for birds. The radar was positioned on the roof of a 12 m building, scanning the sea. Additionally, we deployed an infrared sensor to capture images of the target when the radar detected an object. Fig. 3a shows a flock of seagulls captured in one infrared image. The seagulls had a body size of approximately 0.48 m, and the mean length between adjacent birds was approximately 1.2 m. Using the image scale, we estimate that the length of the flock in the infrared image was approximately 8 m.

FWM signals, resulting from the flapping wings of birds in flight, can be observed in radar echoes of bird flocks. Fig. 3b presents the radar signals of birds in the detection range of approximately 10 km. The values on the vertical axis are dimensionless, while those on the horizontal axis are expressed in units. Sea clutter Doppler frequencies are typically around 0. The time series of the radar signals from the birds were found to be fluctuating. The spectrum displays approximately four clear peaks (-12.9 m/s, -11.4 m/s, -8.4 m/s, -6.3 m/s, and -3.3 m/s), in addition to peaks around 0. The strongest scattering power is generated by the bird's body, observed at a Doppler velocity of -11.4 m/s, while others are produced by the flapping wings of the birds. The FWM effect is responsible for modulating the radar signals from the flapping wings, as described by equations (8, 11, 12). After removing the body Doppler velocity, the radial velocities of the flapping wings were calculated to be +1.5 m/s, -3 m/s, -5.1 m/s, and -8.1 m/s, respectively. Based on the flying direction, it is possible that one bird was flying with a downstroke, while the other two were flapping their wings upward.

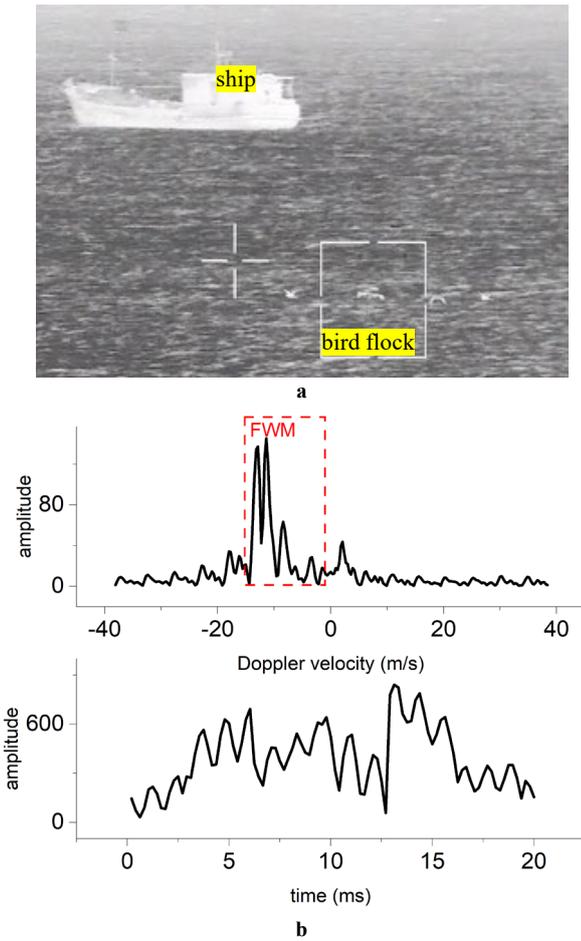

**Fig. 3.** Radar detection of a flock of sea birds. **(a)** A flock of marine birds on an infrared image, **(b)** The FWM signatures in the spectrum and its radar echoes of time series.

Tracking the radial velocities of the flapping wings through the FWM effect reveals the movement patterns of birds in flight. Fig. 4a shows the tracked radial velocities of the flapping wings in a bird flock. After removing the body Doppler velocities, the radial velocities of the flapping wings were calculated. The flock flew from 4.84 km to 8.75 km away from the radar location, into the sea, with an average flight height of approximately 30 m. The tracking continued for 300 s, with an update rate of 1/60 Hz. The mean number of birds during this tracking period was approximately 3.63. However, considering the limited frequency resolution and cross-range bins of the flock length, the number of birds in this flock was rounded up to 4. Additionally, the mean wing-beat frequency of the bird flock, as calculated using formula (15), was approximately 7.0 Hz, consistent with the value measured in the infrared video.

The radar echoes from birds can be modulated by the FWM effect, leading to fluctuations in the radar signals. Fig. 4b shows the tracking of the SNR (signal-to-noise ratio) and SCR (signal-to-clutter ratio) of the birds, both of which were found to be fluctuating. The fluctuation of the SNR values is not in line with the theoretical result, as the SNR value of a target is expected to decrease with increasing radar range according to the radar equation. This suggests that the RCS values of the birds are not constant, and that the flapping wings modulate the RCS values with a fluctuation level of nearly 10 dB, as evidenced by the measured SNR values fluctuating at a level of 10 dB from 2.9 dB to 12.31 dB in Fig. 4b. On the other hand, the SCR values were more stable, fluctuating only between 9.3 dB and 13.8 dB.

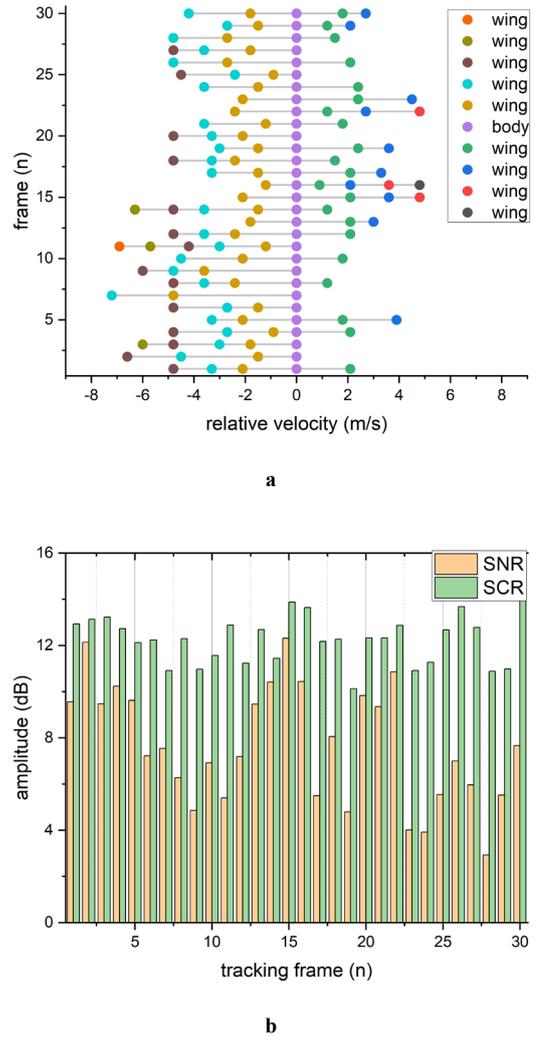

**Fig. 4.** Radar tracking Doppler velocity of flapping wings. **(a)** Velocities of flapping wings after removing body Doppler speed, **(b)** The tracking SNR & SCR values.

## IV. Discussions

The traditional method of quantifying bird numbers based on radar reflectivity may lead to misestimation of bird populations. The current empirical equation for calculating the density of migratory birds using radar reflectivity assumes a constant radar cross-section (RCS) value for birds, given by [12][13][6]:

$$\rho = Z * 28/\sigma \qquad (17)$$

Here, $\rho$ represents bird migration density (unit: Birds/km$^3$), Z represents the radar reflectivity value of the flock (unit: dBZ), and $\sigma$ represents the mean RCS value of a single bird in the bird flock (unit: dBsm). However, recent empirical evidence in radar ornithology has found periodic fluctuations in signal intensities of bird echoes, ranging from 10-20 dB, and in some radar bands, even exceeding 40 dB[2][3][14].These amplitude fluctuations are correlated with the flapping pattern of the bird's wings, indicating that the current assumption of constant bird RCS values in the equation above is a very rough approximation [15][16][17]. Several studies have reported discrepancies in bird migration populations when using equation (17) for quantifying bird numbers based on radar reflectivity in real-world applications. For instance, it has been shown that estimating bird migration density using a mean bird RCS could result in an error of over 100% [18]. Additionally, in some cases, the migration intensity estimated by weather radar did not match with bird radar observations or human observations [17][19][20].

Our Fig. 4b data further supports these findings, showing that using bird RCS values with a 10 dB fluctuation could result in a 100% error in estimated bird migration. In contrast, the FWM effect model can provide a calibration method that accounts for fluctuations caused by flapping wings, leading to a significant improvement in the accuracy of the bird migration quantification model. Moreover, the FWM effect model provides a means of calculating wingbeat rate using radar signals, which may aid in distinguishing between radar signals from birds and non-birds. We plan to further research this topic in the future.

## V. Conclusion

The radar echoes from bird flocks contain modulation signals produced by the flapping wings, resulting in spectral peaks of similar amplitudes spaced at specific intervals, known as the Formation Wing-beat Modulation (FWM) effect. The FWM effect is influenced by the number of birds, their wing-beat frequency, and flight phasing strategy, allowing for accurate quantification of bird numbers in the flock. Additionally, the FWM frequency spacing and Doppler velocities of the flapping wings can be used to calculate the mean wing-beat frequency of the flock. These findings suggest that the FWM effect can be a valuable tool for the radar recognition of bird signals and the quantification of bird behaviors in the fields of radar ornithology and aero-ecology.